\documentclass{ws-jnmp}

\begin{document}

%\arttype{Letter} % default 'Article'

\markboth{LINEARIZATION}
{V.~K.~CHANDRASEKAR, M.~SENTHILVELAN and M.~LAKSHMANAN}

%\catchline{1}{1}{2009}{}{}
%
%%%%%%%%%%%%%%%%%%%%%%%%%%%%%%%%%%%%%%%%%%%%%%%%%%%%%%%%%%%%%%%%%%%%
%\copyrightauthor{D. Henry}

\title{A systematic method of finding linearizing transformations for nonlinear ordinary differential equations: II. Extension to coupled ODEs}

\author{V.~K.~CHANDRASEKAR, M.~SENTHILVELAN and M.~LAKSHMANAN}

\address{Centre for Nonlinear Dynamics, Department of Physics,  Bharathidasan
University, \\ Tiruchirapalli - 620 024, India}
%\email{}

%\author{SECOND AUTHOR}

%\address{Group, Laboratory, Address\\
%City, State ZIP/Zone, Country\\
%\email{author\_id@domain\_name}}

\maketitle

%\begin{history}
%\received{(Day Month Year)}
%\revised{(Day Month Year)}
%\accepted{(Day Month Year)}
%\comby{(xxxxxxxxx)}
%\end{history}

\begin{abstract}
In this second paper on the method of deriving linearizing transformations for nonlinear ODEs, we extend the method to a set of two coupled second order nonlinear ODEs. We show that besides the conventional point, Sundman and generalized linearizing transformations one can also find a large class of mixed or hybrid type linearizing transformations like point-Sundman, point-generalized linearizing transformation and Sundman-generalized linearizing transformation in coupled second order ODEs using the integrals of motion. We propose suitable algorithms to identify all these transformations (with maximal in number) in a  straightforward manner. We illustrate the method of deriving each one of the linearizing transformations with a suitable example.
\end{abstract}
%\pacs{02.30.Hq, 02.30.Ik, 05.45.-a}
%\maketitle
\section{Introduction}
\label{int}
The present paper continues the investigation on the method of finding maximal linearizing transformations of nonlinear ODEs. In the previous paper we have confined our studies to scalar nonlinear ODEs only. In the present paper we extend the method described in Part-I \cite{Chand:11} to the case of two coupled second order nonlinear ODEs. Our results show that in such coupled ODEs there exists a wider class of linearizing transformations including hybrid ones (for example point-Sundman transformation, point-generalized linearizing transformation and Sundman-generalized linearizing transformation).

Unlike the scalar case, the study of linearization of coupled nonlinear ODEs is still in the early stage and only very few results have been established so far. To our knowledge most of the studies are focused on establishing necessary and sufficient conditions, both geometrically and algebraically, for linearization of two coupled second order nonlinear ODEs under invertible point transformations \cite{Fels:95,Crampin:96,Grossman:00,Soh:01,Merker:06,
Qadir:07,Mahomed:07,Sookmee:08,Chand:09a,Mahomed:09,Bagderina:10}.
We note that the necessary and sufficient conditions for the linearization of $N$-coupled second order
nonlinear ODEs under point transformations have been formulated only very recently \cite{Merker:06}.
The formulation of such conditions for transformation other than point transformations is yet to be established. Besides establishing the conditions one also encounters more difficulties in finding explicit linearizing transformations in the case of coupled ODEs since Lie point symmetries and ad-hoc methods rarely help one to get the required transformations for non-point or non-local ones. In certain situations integrations cannot also be performed in a straightforward manner if only a lesser number of integrals are known. Under these circumstances the method proposed here helps one to derive the general solution for the coupled second order nonlinear ODEs besides identifying linearizing transformations.

While extending the algorithm given in Part-I \cite{Chand:11} (for the scalar case) to the case of coupled equations we find that one can have the flexibility of having two types of linearized equations, namely (i) the linearized equations which have the same independent variable and (ii) the linearized equations which have different independent variables. The first type can be captured by (i) point transformation, (ii) Sundman transformation and (iii) generalized linearizing transformation, whereas the linearized equations of the second type can be captured by certain hybrid varieties besides the above mentioned linearizing transformations.

Our studies show that in the first case (the case in which the linearized equations share the same independent variable) one can identify four distinct invertible linearizing point transformations, three Sundman transformations and three generalized linearizing transformations, whereas in the second case one can identify eight invertible point transformations and infinite number of all other transformations from the integrals. To our knowledge, all these results are brought out for the first time in the theory of linearization of nonlinear ODEs.

The plan of the paper is as follows.  In Sec. 2, we present the method of deriving linearizing transformations for the coupled second order ODEs from the first integrals and analyze in-depth the nature of transformations which can be identified through this procedure.  In Sec. 3, we focuss our attention on Type-I linearizing transformations and discuss the
method of finding maximal number of linearizing transformations in each one of the categories, namely
point transformations, Sundman transformations and the generalized linearizing transformations.  We also
illustrate the theory with an example in each category.  In Sec. 4 we discuss the method of finding
maximal number of linearizing transformations of Type II.  Here, we divide our analysis into three groups depending upon the nature of transformations.  For example, we merge the method of finding point-Sundman transformation and point-generalized linearizing transformation into a single category since
these two transformations essentially differ only from the fact that one of the new independent variable
does posses a derivative term or not.  In a similar fashion we merge Sundman transformation, Sundman- generalized linearzing transformation and generalized linearizing transformations into another group and present the method of identifying maximal number of linearizing transformations.  For the sake of illustration, we give separate examples to each one of the linearizing transformations.  Finally, we present our conclusions in Sec. 5.

\section{Extension to coupled ODEs:}
%\label{csode}
\label{sec2:2}
To begin with let us consider two coupled second order nonlinear ODEs of the form
\begin{eqnarray}
\ddot{x}=\phi_1(t,x,y,\dot{x},\dot{y}),\quad \ddot{y}=\phi_2(t,x,y,\dot{x},\dot{y}),
\label{cmet013_eq}
\end{eqnarray}
and assume that Eq. (\ref{cmet013_eq}) admits at least two integrals of the form \cite{Chand:09a}
\begin{eqnarray}
I_1=K_{1}(t,x,y,\dot{x},\dot{y}),\quad I_2=K_{2}(t,x,y,\dot{x},\dot{y}).
\label{cmet013a}
\end{eqnarray}
These two integrals can be constructed using the extended Prelle-Singer procedure \cite{Chand:09}.
As we have shown for the case of scalar second order ODEs, let us recast the integrals in the form
\begin{eqnarray}
I_1&=&\frac{1}{G_1(t,x,y,\dot{x},\dot{y})}\frac{d}{dt}F_{1}(t,x,y)\label{cmet013b}\\
I_2&=&\frac{1}{G_2(t,x,y,\dot{x},\dot{y})}\frac{d}{dt}F_{2}(t,x,y).
\label{cmet013c}
\end{eqnarray}
Now identifying the functions $F_1$, $F_2$ and $G_1$, $G_2$ as new dependent and independent variables,
namely
\begin{eqnarray}
w_1 &= &F_1(t,x,y),\;z_1 = \int_o^t G_1(t',x,y,\dot{x},\dot{y}) dt'\nonumber\\ w_2& = &F_2(t,x,y),\;z_2 = \int_o^t G_2(t',x,y,\dot{x},\dot{y}) dt',
\label{cmet013d}
\end{eqnarray}
equation (\ref{cmet013a}) can be brought to the form
\begin{eqnarray}
\hat{I}_i=dw_i/dz_i,  \;\;\;\; i=1,2,
\label{cmet013e}
\end{eqnarray}
which in turn provides the necessary two independent free particle equations, namely $d^2w_i/dz_i^2=0$, $i=1,2$, upon differentiation.

\subsection{The nature of transformations \cite{Chand:09a}}
\label{sec2:3}
Unlike the scalar case, presently we have two independent variables, namely $z_1$ and $z_2$. As a consequence we have the flexibility of fixing them either as the same variable or as different variables, that is (i) $z_1=z_2$ (Type I) or (ii) $z_1\neq z_2$ (Type II). In the first case one can deduce point transformation, Sundman transformation and generalized linearizing
transformation. In all these cases the new independent variable is the same in both the equations. On the other hand relaxing the condition,  that is, the
new independent variables need not be the same in the linearized free particle equations, one gets $d^2w_i/dz_i^2=0$, $i=1,2$, which allows one to identify a larger class of linearizing transformations, as we see below.

\subsubsection{Type-I Linearizing Transformations ($z_1=z_2=z$)}
\label{sec2:3:1} In the case of Type-I transformations we have $w_1 = F_{1}(t,x,y),
\;w_2 = F_{2}(t,x,y)$, $z_1 =z_2 =z= \int
G_1(t, x, y,\dot{x}, \dot{y}) dt = \int
G_2(t, x, y,$ $ \dot{x}, \dot{y}) dt$. Now appropriately restricting the
form of $G_1$ $(=G_2)$, one can identify three different types of linearizing
transformations.
\begin{enumerate}
\item{
Suppose $z_1=z_2=z$ is a perfect
differential function and $w_i$'s, $i=1,2,$ and $z$ do not
contain the variables $\dot{x}$ and $\dot{y}$, then we call the
resultant transformation, namely $w_1=F_1(t,x,y)$, $w_2=F_2(t,x,y)$,
$z=\hat{G}(t,x,y)$, as a point transformation of type-I.}
\item{ On
the other hand, if $z$ is not a perfect differential
function, and  $w_i$'s, $i=1,2,$ and $z$ do not contain the
variables $\dot{x}$ and $\dot{y}$, then we call the resultant
transformation, $w_1=F_1(t,x,y)$, $w_2=F_2(t,x,y)$,
$z=\int G(t,x,y)dt$, as a Sundman transformation
of type-I.}
\item{ As a more general case, if we consider the independent
variable $z$ to
contain the derivative terms also, that is $w_1=F_1(t,x,y)$,
$w_2=F_2(t,x,y)$, $z=\int G(t,x,y,\dot{x},\dot{y})dt$, we have the resultant transformation as a generalized linearizing
transformation of type-I.}
\end{enumerate}
%In our analysis, we do not consider the possibility $w_1=F_1(t,x,y,\dot{x},\dot{y})$,
%$w_2=F_2(t,x,y,\dot{x},\dot{y})$ and $z=G(t,x,y,\dot{x},\dot{y})$ because the
%procedure to handle such a transformation is different from the presently discussed
%linearizing transformations. This possibility will be studied separately.
\subsubsection{Type-II Linearizing Transformations ($z_1\neq z_2$)}
\label{sec2:3:2} In the Type-II linearizing transformations we have $w_1 =
F_{1}(t,x,y), \;w_2 = F_{2}(t,x,y)$ and $z_1 =\int
G_1(t,x,y,\dot{x},\dot{y}) dt,$ $z_2 = \int
G_2(t,x,y,\dot{x},\dot{y}) dt$, $z_1\ne z_2$.  Now appropriately
restricting the
forms of $G_1$ and $G_2$, one can get six  different types of
linearizing transformations.
\begin{enumerate}
\item{ If $z_1$ and $z_2$
are perfect differential functions and $w_i$'s and $z_i$'s,
$i=1,2,$ do not contain the variables $\dot{x}$ and $\dot{y}$, then we call the resultant transformation, namely $w_1=F_1(t,x,y)$,
$w_2=F_2(t,x,y)$, $z_1=G_1(t,x,y)$, $z_2=G_2(t,x,y)$, as a point
transformation of type-II.}
\item{ Suppose $z_1$ is a perfect differential
function and $z_2$ is not a perfect differential function or
vice versa, and if $z_1$
and $z_2$ do not contain the variables $\dot{x}$ and $\dot{y}$, then
we can call the resultant transformation, namely $w_1=F_1(t,x,y)$,
$w_2=F_2(t,x,y),\;z_1=G_1(t,x,y)$, $z_2=\int G_2(t,x,y)dt$ or
$z_1=\int G_1(t,x,y)dt$, $z_2=G_2(t,x,y)$, as a mixed
point-Sundman transformation.}
\item{On the other hand, if any one of the independent variables
contains the variables $\dot{x}$ and $\dot{y}$, we call the resultant
transformation, namely $w_1=F_1(t,x,y)$,
$w_2=F_2(t,x,y),\;z_1=G_1(t,x,y)$, $z_2=\int
G_2(t,x,y,\dot{x},\dot{y})dt$ or $z_1=\int
G_1(t,x,y,\dot{x},\dot{y})dt$ and  $z_2=G_2(t,x,y)$, as a
mixed point-generalized linearizing transformation.}
\item{ Suppose the independent variables
are not perfect differential functions and are also not functions of
$\dot{x}$ and $\dot{y}$, that is, $w_1=F_1(t,x,y)$,
$w_2=F_2(t,x,y),\;z_1=\int G_1(t,x,y)dt$, $z_2=\int G_2(t,x,y)dt$,
then we call the resultant transformation as a Sundman
transformation of type-II.}
\item{ Further, if one of the independent variables, say $z_1$, does not contain the
derivative terms while the other independent variable $z_2$ does
contain the derivative terms or
vice versa, that is
$w_1=F_1(t,x,y)$, $w_2=F_2(t,x,y)$, $z_1=\int G_1(t,x,y)dt$,
$z_2=\int G_2(t,x,y,\dot{x},\dot{y})dt$ or $z_1=\int
G_1(t,x,y,\dot{x},\dot{y})dt$, $z_2=\int G_2(t,$ $x,y)dt$, then we call
the resultant
transformation as a mixed Sundman-generalized
linearizing transformation.}
\item{ As a general case, if we allow both the independent variables,
$z_1$ and $z_2$, to be non-perfect
differential functions and also to contain derivative terms,  that is,
$w_1=F_1(t,x,y)$,  $w_2=F_2(t,x,y)$, $z_1=\int
G_1(t,x,y,\dot{x},\dot{y})dt$, $z_2=\int
G_2(t,x,y,\dot{x},\dot{y})dt$, then the resultant
transformation will be termed as a generalized linearizing transformation of
type-II.}
\end{enumerate}

In the following we will discuss the method of deriving maximal number of linearizing transformations for each one of the above cases.

\section{Type-I Linearizing Transformations ($z_1=z_2=z$)}
\label{sec3}
\subsection{Invertible Point Transformation}
\label{sec3:1}
\subsubsection{First set}
\label{sec3:1:1}
In this case $w_1 = F_1(t,x,y)$, $w_2 = F_2(t,x,y)$ and
$z = \int_o^t G(t',x,y,\dot{x},\dot{y}) dt'=\hat{G}(t,x,y)$ is the first pair of linearizing point transformations for the given equation which can be readily identified from the two integrals, see Sec.
\ref{sec2:3:1} above.

Integrating the free particle equations we obtain the general solution of the form
\begin{eqnarray}
 w_1=I_1z+I_3,\quad w_2=I_2z+I_4,\label{cmet04}
\end{eqnarray}
where $I_1, I_2, I_3$ and $I_4$ are four integration constants. Now rewriting $w_1,\;w_2$ and $z$ in terms of the old variables one obtains the general solution for the given system of two coupled second order nonlinear ODEs.

\subsubsection{Second set}
\label{sec3:1:2}
We identify the second set of linearizing transformations from the integrals $I_3$ and $I_4$. To do so we  utilize the expressions (\ref{cmet04}) to express $I_3$ and $I_4$ and derive the remaining LTs.

Rewriting (\ref{cmet04}) in the form
\begin{eqnarray}
I_3&=w_1-I_1z&=\displaystyle{w_1-\frac {dw_1}{dz}z},\nonumber\\
I_4&=w_2-I_2z&=w_2-\frac {dw_2}{dz}z,\label{cmet06}
\end{eqnarray}
and replacing $w_1,\;w_2$ and $z$ in terms of $F_{1},\;F_2$ and $\hat{G}$, respectively, and $\frac{dw_1}{dz}$ and $\frac{dw_2}{dz}$ as $\frac{\dot{F_{1}}}{\dot{\hat{G}}}$ and $\frac{\dot{F_{2}}}{\dot{\hat{G}}}$, respectively, we get
\begin{eqnarray}
 I_3=\frac{1}{\dot{\hat{G}}}\bigg(F_{1}\dot{\hat{G}}-\dot{F_{1}}\hat{G}\bigg),\quad
I_4=\frac{1}{\dot{\hat{G}}}\bigg(F_{2}\dot{\hat{G}}-\dot{F_{2}}\hat{G}\bigg).\label{cmet08}
\end{eqnarray}
Now let us split the expressions which appear on the right hand side in (\ref{cmet08}) further into two
perfect derivative functions such that they will provide us the second set of
linearizing transformations.

To start with we choose the dependent variables by rewriting (\ref{cmet08}) in the form
\begin{eqnarray}
 I_3=-\frac{(\hat{G})^2}{\dot{\hat{G}}}
\frac{d}{dt}\bigg(\frac{F_{1}}{\hat{G}}\bigg),\quad
I_4=-\frac{(\hat{G})^2}{\dot{\hat{G}}}
\frac{d}{dt}\bigg(\frac{F_{2}}{\hat{G}}\bigg).\label{cmet09a}
\end{eqnarray}
The above forms can be re-expressed as
\begin{eqnarray}
I_3=\frac{1}{\displaystyle{\frac{d}{dt}\bigg(\frac{1}{\hat{G}}\bigg)}}
\frac{d}{dt}\bigg(\frac{F_{1}}{\hat{G}}\bigg),\quad
I_4=\frac{1}{\displaystyle{\frac{d}{dt}\bigg(\frac{1}{\hat{G}}\bigg)}}
\frac{d}{dt}\bigg(\frac{F_{2}}{\hat{G}}\bigg).\label{cmet09b}
\end{eqnarray}
Now identifying
\begin{eqnarray}
 w_{11} = \frac{F_{1}}{\hat{G}},\quad w_{21}= \frac{F_{2}}{\hat{G}},\quad
z_1 = \frac{1}{\hat{G}}, \label{cmet10}
\end{eqnarray}
equation (\ref{cmet09b}) can be brought to the form
\begin{eqnarray}
I_3=\frac {dw_{11}}{dz_1},\quad I_4=\frac {dw_{21}}{dz_1}.
\label{cmet10a1}
\end{eqnarray}
In other words we have
\begin{eqnarray}
\frac {d^2w_{11}}{dz_1^2}=0, \quad \frac {d^2w_{21}}{dz_1^2}=0.
\end{eqnarray}
Through this procedure one can get a second set of linearizing point transformations from the integrals
$I_3$ and $I_4$.
\subsubsection{Third set}
We identify a third set of linearizing transformations by rewriting the expression (\ref{cmet08}) in the form
\begin{eqnarray}
I_3=\frac{\dot{\hat{G}}}{\dot{F_{1}}}\bigg(\frac{F_{1}\dot{F_{1}}}{\dot{\hat{G}}}
-\frac{\dot{F_{1}}^2}{\dot{\hat{G}}^2}\hat{G} \bigg),
\quad I_4=\frac{1}{\dot{F_{1}}}\bigg(F_{2}\dot{F_{1}}+I_3\dot{F_{2}}-F_{1}\dot{F_{2}}\bigg).
\label{cmet11}
\end{eqnarray}
Using the identity $I_1=\frac{\dot{F_{1}}}{\dot{\hat{G}}}$ and $I_2=\frac{\dot{F_{2}}}{\dot{\hat{G}}}$,
(vide Eqs.(\ref{cmet013b}) and (\ref{cmet013c})) the above equation (\ref{cmet11}) can be simplified to
\begin{eqnarray}
 I_3=\frac{I_1}{\dot{F_{1}}}\bigg(F_{1}\dot{\hat{G}}-\dot{F_{1}}\hat{G} \bigg),\quad
I_4-\frac{I_3I_2}{I_1}=\frac{1}{\dot{F_{1}}}\bigg(F_{2}\dot{F_{1}}-F_{1}\dot{F_{2}}\bigg).
\label{cmet12}
\end{eqnarray}
Interestingly the right hand sides of both the equations in (\ref{cmet12}) can be brought to the forms of total derivatives of suitable functions. To do so let us rewrite the right hand sides of the above two expressions in the form
\begin{eqnarray}
 \hat{I_3}=\frac{F_{1}^2}{\dot{F_{1}}}
\frac{d}{dt}\bigg(\frac{\hat{G}}{F_{1}}\bigg),\quad
\hat{I_4}=\frac{F_{1}^2}{\dot{F_{1}}}
\frac{d}{dt}\bigg(\frac{-F_{2}}{F_{1}}\bigg),\label{cmet13a}
\end{eqnarray}
where $\hat{I_3}=\frac{I_3}{I_1}$ and $\hat{I_4}=\frac{I_1I_4-I_2I_3}{I_1}$.
Proceeding further we find that the first term can be recast in the form
\begin{eqnarray}
\hat{I_3}=\frac{1}{\frac{d}{dt}(-\frac{1}{F_{1}})}
\frac{d}{dt}\bigg(\frac{\hat{G}}{F_{1}}\bigg), \quad
\hat{I_4}=\frac{1}{\frac{d}{dt}(-\frac{1}{F_{1}})}
\frac{d}{dt}\bigg(\frac{-F_{2}}{F_{1}}\bigg).
\label{cmet13b}
\end{eqnarray}
Now choosing
\begin{eqnarray}
 w_{12} = \frac{\hat{G}}{F_{1}},\quad w_{22} =- \frac{F_{2}}{F_{1}},
\quad z_2 = -\frac{1}{F_{1}},\label{cmet14}
\end{eqnarray}
equation  (\ref{cmet13b}) becomes
\begin{eqnarray}
\hat{I_3}= \frac{dw_{12}}{dz_2}, \quad \hat{I_4}= \frac{dw_{22}}{dz_2}.
\label{cmet14a1}
\end{eqnarray}

Thus, the set $w_{12},\;w_{22}$ and $z_2$ becomes the third pair of linearizing point transformations for the given set of two coupled second order nonlinear ODEs.
\subsubsection{Fourth set}
We observe that one may rewrite (\ref{cmet08}) also in the form
\begin{eqnarray}
I_3=\frac{1}{\dot{F_{2}}}\bigg(F_{1}\dot{F_{2}}+I_4\dot{F_{1}}-F_{2}\dot{F_{1}}\bigg),
\quad I_4=\frac{G}{\dot{F_{2}}}\bigg(\frac{F_{2}\dot{F_{2}}}{G}
-\frac{\dot{F_{2}}^2}{\dot{\hat{G}}^2}\hat{G} \bigg).
\label{cmet15}
\end{eqnarray}
Using again the identity $I_1=\frac{\dot{F_{1}}}{\dot{\hat{G}}}$ and $I_2=\frac{\dot{F_{2}}}{\dot{\hat{G}}}$,
equation (\ref{cmet15}) can be simplified to
\begin{eqnarray}
I_3-\frac{I_4I_1}{I_2}=\frac{1}{\dot{F_{2}}}\bigg(\dot{F_{2}}F_{1}-\dot{F_{1}}F_{2}\bigg),\quad
I_4=\frac{I_2}{\dot{F_{2}}}\bigg(F_{2}G-\dot{F_{2}}\hat{G}\bigg).
\label{cmet16}
\end{eqnarray}
Interestingly the right hand sides of both the expressions in
(\ref{cmet16}) can also be rewritten as a product of two perfect derivative terms as we see below.

Rewriting the terms inside the bracket  in the right hand side expressions of (\ref{cmet16}) as perfect derivatives, that is
\begin{eqnarray}
 \bar{I_3}=\frac{F_{2}^2}{\dot{F_{2}}}
\frac{d}{dt}\bigg(\frac{-F_{1}}{F_{2}}\bigg),\quad
\bar{I_4}=\frac{F_{2}^2}{\dot{F_{2}}}
\frac{d}{dt}\bigg(\frac{\hat{G}}{F_{2}}\bigg),\label{cmet17a}
\end{eqnarray}
where $\bar{I_3}=\frac{I_2I_3-I_1I_4}{I_2}$ and $\bar{I_4}=\frac{I_4}{I_2}$, and rewriting the prefactors suitably, we arrive at
\begin{eqnarray}
 \hat{I_3}=\frac{1}{\frac{d}{dt}(-\frac{1}{F_{2}})}
\frac{d}{dt}\bigg(\frac{-F_{1}}{F_{2}}\bigg), \quad
\hat{I_4}=\frac{1}{\frac{d}{dt}(-\frac{1}{F_{2}})}
\frac{d}{dt}\bigg(\frac{\hat{G}}{F_{2}}\bigg).
\label{cmet17b}
\end{eqnarray}
Now choosing the new dependent and independent variables in the following way,
\begin{eqnarray}
 w_{13} = \frac{-F_{1}}{F_{2}},\quad w_{23} = \frac{\hat{G}}{F_{2}},
\quad z_3 = \frac{-1}{F_{2}},\label{cmet18}
\end{eqnarray}
equation (\ref{cmet17b}) can be brought to the form
\begin{eqnarray}
\bar{I_3}= \frac{d w_{13}}{dz_3}, \quad \bar{I_4}= \frac{d w_{23}}{dz_3}.
\label{cmet18a1}
\end{eqnarray}
As a consequence one can arrive at the free particle equation through the fourth set of variables, namely $ w_{13},\; w_{23}$ and $z_3$.

Summarizing, we find that one can deduce four types of linearizing point transformations from the integrals for a linearizable two coupled second order nonlinear ODE, namely

(i) $w_1 = F_1(t,x,y)$, $w_2 = F_2(t,x,y)$ and
$z = \int_o^t G(t',x,y,\dot{x},\dot{y}) dt'=\hat{G}(t,x,y)$,

(ii) $w_{11} = \frac{F_{1}}{\hat{G}}$, $w_{21}= \frac{F_{2}}{\hat{G}}$ and $z_1 = \frac{1}{\hat{G}}$,

(iii) $w_{12} = \frac{\hat{G}}{F_{1}}$, $w_{22} =- \frac{F_{2}}{F_{1}}$ and $z_2 = -\frac{1}{F_{1}}$ and

(iv) $w_{13} = \frac{-F_{1}}{F_{2}}$, $w_{23} = \frac{\hat{G}}{F_{2}}$ and $z_3 = \frac{-1}{F_{2}}$.

One may observe that the first three sets of transformations are similar to the scalar case (with appropriate extension to the second dependent variable) and the fourth set is similar to the third set with suitable changes in the new dependent and independent variables.

\subsubsection{Transformation by interchange of variables}
\label{tranint}
Finally, we note that since the dependent and independent variables in the point transformations are of invertible type, one can enumerate few more linearizing point transformations just by interchanging the dependent and independent variables in the expressions (\ref{cmet10a1}), (\ref{cmet14a1}) and (\ref{cmet18a1}). In fact, the set of transformations, (i) $w_1 = \hat{G}$, $w_2 = F_2$ and $z = F_1$, (ii) $w_1 = F_1$, $w_2 = \hat{G}$ and $z = F_2$, (iii) $w_{11} =  \frac{1}{\hat{G}}$, $w_{21}= \frac{F_{2}}{\hat{G}}$ and $z_1 =  \frac{F_{1}}{\hat{G}}$, (iv) $w_{11} = \frac{F_{1}}{\hat{G}}$, $w_{21}=  \frac{1}{\hat{G}} $ and $z_1 = \frac{F_{2}}{\hat{G}}$, (v) $w_{12} = -\frac{1}{F_{1}}$, $w_{22} =- \frac{F_{2}}{F_{1}}$
and $z_2 = \frac{\hat{G}}{F_{1}}$, (vi) $w_{12} = \frac{\hat{G}}{F_{1}}$, $w_{22} =-\frac{1}{F_{1}}$
and $z_2 = - \frac{F_{2}}{F_{1}}$, (vii) $w_{13} = \frac{-1}{F_{2}}$, $w_{23} = \frac{\hat{G}}{F_{2}}$
and $z_3 = \frac{-F_{1}}{F_{2}}$ and (viii) $w_{13} = \frac{-F_{1}}{F_{2}}$, $w_{23} = \frac{-1}{F_{2}}$ and $z_3 =\frac{\hat{G}}{F_{2}}$, also form a nontrivial set of linearizing point transformations for the given two coupled second order nonlinear ODEs. These eight transformations can be identified from the above four pairs by interchanging the independent variable in place of a dependent variable and vice versa.

\subsubsection{Transformation through linear combination}
\label{linear com}
We also note that in the case of type-I linearizing transformations ($z_1=z_2=z$), $w_1$ and $w_2$ act as two new dependent variables. Then the combination $u_1=a_1w_1+a_2w_2$ and $u_2=b_1w_1+b_2w_2$, where $a_1$, $a_2$, $a_3$ and $b_2$ are arbitrary scalar constants, also acts as a set of two new dependent variables. The new independent variable is $z$.

\subsubsection{Nonexistence of other sets}
Deducing point transformation from the integrals of motion is possible if and only if the integrals are ratios of two polynomials where each polynomial is linear in the first derivative of the dependant variables. To find out all the possible combinations of integrals that are linear in the first derivatives let us consider a more general form of the integral which is given by

\begin{eqnarray}
I&=&\sum_{j=1}^{4}s_jI_{1}^{r_{1j}}I_2^{r_{2j}}
I_3^{r_{3j}}I_4^{r_{4j}}\nonumber\\
&=&\sum_{j=1}^{4}s_j\frac{\dot{F_1}^{r_{1j}}\dot{F_2}^{r_{2j}}}{\dot{\hat{G}}^{r_{1j}+r_{2j}+r_{3j}+r_{4j}}}
\bigg(F_{1}\dot{\hat{G}}-\dot{F_{1}}\hat{G}\bigg)^{r_{3j}}
\bigg(F_{2}\dot{\hat{G}}-\dot{F_{2}}\hat{G}\bigg)^{r_{4j}},
\label{more01}
\end{eqnarray}

where $s_i$'s and $r_{ij}$'s, $i,j=1,2,3,4$, are some real numbers. The possible linear combination of integrals that provide the linearizing transformations can be collected in the following way:
\begin{eqnarray}
K_{(1,2)}^{(1)}&=&a_{(1,2)}I_1+b_{(1,2)}I_2+c_{(1,2)}I_3+d_{(1,2)}I_4\nonumber\\
&=&\frac{1}{{\dot{\hat{G}}}}\bigg[a_{(1,2)}\dot{F_1}+b_{(1,2)}\dot{F_2}+
c_{(1,2)}\bigg(F_{1}\dot{\hat{G}}-\dot{F_{1}}\hat{G}\bigg)+
d_{(1,2)}\bigg(F_{2}\dot{\hat{G}}-\dot{F_{2}}\hat{G}\bigg)\bigg],\nonumber\\
K_{(1,2)}^{(2)}&=&a_{(1,2)}\frac{1}{I_1}+b_{(1,2)}\frac{I_2}{I_1}+c_{(1,2)}\frac{I_3}{I_1}
+d_{(1,2)}\frac{I_4}{I_1}+e_{(1,2)}(I_4-\frac{I_2I_3}{I_1})
\nonumber\\
&=&\frac{1}{\dot{F_1}}\bigg[a_{(1,2)}\dot{\hat{G}}+b_{(1,2)}\dot{F_2}
+c_{(1,2)}\bigg(F_{1}\dot{\hat{G}}-\dot{F_{1}}\hat{G}\bigg)+
d_{(1,2)}\bigg(F_{2}\dot{\hat{G}}-\dot{F_{2}}\hat{G}\bigg)
\nonumber\\&&
+e_{(1,2)}\bigg(F_{2}\dot{F_{1}}-F_{1}\dot{F_{2}}\bigg)\bigg],
\nonumber\\
K_{(1,2)}^{(3)}&=&a_{(1,2)}\frac{I_1}{I_2}+b_{(1,2)}\frac{1}{I_2}+c_{(1,2)}\frac{I_3}{I_2}
+d_{(1,2)}\frac{I_4}{I_2}+e_{(1,2)}(I_3-\frac{I_1I_4}{I_2})
\nonumber\\
&=&\frac{1}{\dot{F_2}}\bigg[a_{(1,2)}\dot{F_1}+b_{(1,2)}\dot{\hat{G}}+
c_{(1,2)}\bigg(F_{1}\dot{\hat{G}}-\dot{F_{1}}\hat{G}\bigg)+
d_{(1,2)}\bigg(F_{2}\dot{\hat{G}}-\dot{F_{2}}\hat{G}\bigg)
\nonumber\\&&
+e_{(1,2)}\bigg(F_{1}\dot{F_{2}}-F_{2}\dot{F_{1}}\bigg)\bigg],\nonumber
\end{eqnarray}

\begin{eqnarray}
K_{(1,2)}^{(4)}&=&a_{(1,2)}\frac{I_1}{I_3}+b_{(1,2)}\frac{I_2}{I_3}+c_{(1,2)}\frac{1}{I_3}
+d_{(1,2)}\frac{I_4}{I_3}+e_{(1,2)}(I_2-\frac{I_1I_4}{I_3})\nonumber\\
&=&\frac{1}{\bigg(F_{1}\dot{\hat{G}}-\dot{F_{1}}\hat{G}\bigg)}\bigg[a_{(1,2)}\dot{F_1}+b_{(1,2)}\dot{F_2}+
c_{(1,2)}\dot{\hat{G}}+
d_{(1,2)}\bigg(F_{2}\dot{\hat{G}}-\dot{F_{2}}\hat{G}\bigg)
\nonumber\\&&
+e_{(1,2)}\bigg(F_{1}\dot{F_{2}}-F_{2}\dot{F_{1}}\bigg)\bigg],\nonumber\\
K_{(1,2)}^{(5)}&=&a_{(1,2)}\frac{I_1}{I_4}+b_{(1,2)}\frac{I_2}{I_4}+c_{(1,2)}\frac{I_3}{I_4}
+d_{(1,2)}\frac{1}{I_4}+e_{(1,2)}(I_1-\frac{I_3I_2}{I_4})\nonumber\\
&=&\frac{1}{\bigg(F_{2}\dot{\hat{G}}-\dot{F_{2}}\hat{G}\bigg)}\bigg[a_{(1,2)}\dot{F_1}+b_{(1,2)}\dot{F_2}+
c_{(1,2)}\bigg(F_{1}\dot{\hat{G}}-\dot{F_{1}}\hat{G}\bigg)+
d_{(1,2)}\dot{\hat{G}}
\nonumber\\&&
+e_{(1,2)}\bigg(F_{1}\dot{F_{2}}-F_{2}\dot{F_{1}}\bigg)\bigg],\nonumber\\
K_{(1,2)}^{(6)}&=&\frac{1}{I_1I_4-I_3I_2}(a_{(1,2)}I_1+b_{(1,2)}I_2+c_{(1,2)}I_3
+d_{(1,2)}I_4)\nonumber\\
&=&\frac{1}{\bigg(F_{1}\dot{F_{2}}-F_{2}\dot{F_{1}}\bigg)}\bigg[a_{(1,2)}\dot{F_1}+b_{(1,2)}\dot{F_2}+
c_{(1,2)}\bigg(F_{1}\dot{\hat{G}}-\dot{F_{1}}\hat{G}\bigg)
\nonumber\\&&+
d_{(1,2)}\bigg(F_{2}\dot{\hat{G}}-\dot{F_{2}}\hat{G}\bigg)\bigg],
\label{more03}
\end{eqnarray}
where $a_{(1,2)}$, $b_{(1,2)}$, $c_{(1,2)}$, $d_{(1,2)}$ and $e_{(1,2)}$ are arbitrary real numbers. Here $K_{(1)}^{(i)}$ and $K_{(2)}^{(i)}$, $i=1, 2, \ldots, 6$, are the first and second integrals of the $i$th collection. For the linearizing point transformations of type I we find that all the four sets found in Sec. \ref{sec3:1} can be extracted from the above. Any other set which can be extracted from the above other than the basic four sets are either found to be a linear combination of the dependent variables or an interchange of the dependent and the independent variable within a set. Thus we conclude that there exist only four basic sets of invertible linearizing point transformations of type-I for a given coupled second order ODE.

\subsubsection{Example:1 }
Now we illustrate the procedure proposed above by considering an example \cite{Soh:01}
\begin{eqnarray}
\ddot{x}=\dot{x}^2+\dot{y}^2,\quad \ddot{y}=2\dot{x}\dot{y}. \label{fcou10}
\end{eqnarray}
The first integrals, which can be obtained using the formulation given in Chandrasekar et al \cite{Chand:09}, can be
written as
\begin{eqnarray}
I_1 =-(\dot{x}+\dot{y})e^{-(x+y)}, \quad
I_2=(\dot{y}-\dot{x})e^{(y-x)}. \label{cou10}
\end{eqnarray}
Rewriting (\ref{cou10}), we get
\begin{eqnarray}
I_1& =\displaystyle{\frac {d}{dt}e^{-(x+y)}}
= \frac {dw_1}{dz_1},\quad
I_2& =\displaystyle{\frac {d}{dt}e^{(y-x)}}= \frac {dw_2}{dz_2}\label{cou11b}
\end{eqnarray}
so that
\begin{eqnarray}
w_1=e^{-(x+y)},\quad w_2=e^{(y-x)},\quad z=t. \label{cou12}
\end{eqnarray}
Since the transformation is of invertible point type one can also express old
coordinates in terms of new coordinates in the form
\begin{eqnarray}
x=-\frac{1}{2}\log(w_1w_2),\quad y=\frac{1}{2}\log(\frac{w_2}{w_1}),\quad
t=z.\label{cou12a}
\end{eqnarray}
By utilizing the transformation (\ref{cou12a}), one can transform
(\ref{fcou10}) to the second order free
particle equations, namely, $\frac{d^2w_1}{dz^2}=0$ and
$\frac{d^2w_2}{dz^2}=0$.

We get the remaining two integrals of motion for equation (\ref{fcou10}) in the form \cite{Chand:09},
\begin{eqnarray}
I_3 =(1+(\dot{x}+\dot{y})t)e^{-(x+y)}, \quad
I_4=(1+(\dot{x}-\dot{y})t)e^{(y-x)}. \label{2scou10}
\end{eqnarray}
Let us rewrite (\ref{2scou10}) as
\begin{eqnarray}
I_3& =\displaystyle{-t^2\frac {d}{dt}(\frac{e^{-(x+y)}}{t})}
= \frac {d\hat{w}_1}{d\hat{z}},\label{2scou11a}\\
I_4& =\displaystyle{-t^2
\frac {d}{dt}(\frac{e^{(y-x)}}{t})}= \frac {d\hat{w}_2}{d\hat{z}},\label{2scou11b}
\end{eqnarray}
where
\begin{eqnarray}
w_{11}=\frac{e^{-(x+y)}}{t},\quad w_{21}=\frac{e^{(y-x)}}{t},\quad z_1=\frac{1}{t}. \label{2scou12}
\end{eqnarray}
Using the new variables~(\ref{2scou12}) one can transform (\ref{fcou10}) to the two uncoupled free
particle equations.

The other two linearizing point transformations can be obtained straightforwardly from the expressions given in (\ref{cmet14}) and (\ref{cmet18}) which in turn read
\begin{eqnarray}
w_{12}=te^{(x+y)},\quad w_{22}=e^{2y},\quad z_2=e^{(x+y)}.
\label{2scou14}
\end{eqnarray}
\begin{eqnarray}
w_{13}=e^{-2y},\quad w_{23}=te^{(x-y)},\quad z_3=e^{(x-y)}. \label{2scou15}
\end{eqnarray}
Besides the above, by interchanging the dependent and independent variables in the expressions given in Sec. \ref{tranint} and through their linear combinations (Sec. \ref{linear com}) one can obtain more sets of linearizing transformations additionally for the Eq. (\ref{fcou10}).

\subsection{Sundman and generalized linearizing transformations of type-I(GST-I and GLT-I)}
In Sec. \ref{sec2:2} we noted that GST-I differs from GLT-I only in the independent variable part. As a consequence the method of identifying the linearizing transformations from the integrals for both the types can be grouped together.
\subsubsection{First pair of transformation}
The first pair of linearizing transformations for both the cases can be readily identified by writing the first and second integrals in the form $I_1=\frac{1}{\frac{dz}{dt}}\frac{dF_1}{dt}=\frac{dw_1}{dz}$ and $I_2=\frac{1}{\frac{dz}{dt}}\frac{dF_2}{dt}=\frac{dw_2}{dz}$. The resultant transformation reads
\begin{eqnarray}
\mbox{GST}: & w_1=F_1(x,y,t),\quad w_2=F_2(x,y,t),&\quad dz=G(t,x,y)dt,\nonumber\\
\mbox{GLT}: &w_1=F_1(x,y,t),\quad w_2=F_2(x,y,t),&\quad dz=G(t,x,y,\dot{x},\dot{y})dt,
\label {ceq06}
\end{eqnarray}
respectively.

\subsubsection{Second and third pairs}
To derive the additional pair of linearizing transformations we recall the expression $w_1=I_1z+C$, where $C$ is a constant, (which is obtained by integrating the first of the linearized equation in (\ref{cmet013e})) and use the observation that $\frac{I_1}{I_2}$ is a perfect derivative, that is (since $G_1=G_2$)
\begin{eqnarray}
\frac{I_1}{I_2}=\frac{1}{\frac{dF_2}{dt}}\frac{dF_1}{dt}=\frac{dw_1}{dw_2}.
\end{eqnarray}
From the above we can fix the third integral $I_3$ in the form
\begin{eqnarray}
dw_1= \frac{I_1}{I_2}dw_2\quad \Rightarrow \quad w_1=\frac{I_1}{I_2}w_2+I_3.
\label{cspt01ba}
\end{eqnarray}

Once an expression for the third integral is obtained the latter can be rewritten suitably in order to yield the second set of linearizing transformations for the given equation. In the following we illustrate this procedure.

Rewriting (\ref{cspt01ba}) in the form
\begin{eqnarray}
I_3=w_1-\frac{I_1}{I_2}w_2=\displaystyle{w_1-\frac {dw_1}{dw_2}w_2}\label{spgst01}
\end{eqnarray}
and replacing $w_1$, $w_2$ and $\frac{dw_1}{dw_2}$ in terms of $F_1$ and $F_2$,
we get
\begin{eqnarray}
I_3&=&
\displaystyle{F_1-\frac{1}{\frac{dF_2}{dt}}\frac{dF_1}{dt}F_2},
\label{spgst02}\nonumber\\
&=&\displaystyle{\frac{1}{\dot{F_2}}\bigg(F_1\dot{F_2}-\dot{F_1}F_2\bigg)}.
\label{spgst03}
\end{eqnarray}
Eq. (\ref{spgst03}) can be recast in the form
\begin{eqnarray}
I_3=-\displaystyle{\frac{1}{\frac{d}{dt}\bigg(\frac{1}{F_2}\bigg)}}
\frac{d}{dt}\bigg(\frac{F_1}{F_2}\bigg).\label{spgst05}
\end{eqnarray}
We can immediately identify the new variables as
\begin{eqnarray}
w_{1} = \frac{F_1}{F_2},\quad
w_2 = \frac{1}{F_2},\quad dz=\frac{G}{F_2^2}dt \label{spgst06}
\end{eqnarray}
which in turn constitutes the second set of linearizing transformations.

On the other hand starting from $w_2=I_2z$ and the identity $\frac{I_2}{I_1}=\frac{dw_2}{dw_1}$ one can get $w_2=\frac{I_2}{I_1}w_1+\hat{I_3}$. Following the procedure given above one can identify the third set of linearizing transformations in the form
\begin{eqnarray}
w_1=\frac{1}{F_1},\quad w_2=\frac{F_2}{F_1},\quad dz=\frac{G}{F_1^2}dt.
\label {ceq15}
\end{eqnarray}
The result reveals the fact that for both the type-I GST ($G$ is a function of $t$, $x$, $y$) and GLT (whenever $G$ is a function of $t$, $x$, $y$, $\dot{x}$, $\dot{y}$) one can identify three sets of linearizing transformations from the integrals. We note here that unlike the scalar case we do not have the privilege of having infinite sets of linearizing transformations. This can be illustrated as follows:  As
we did in the scalar case let us rewrite the integral $I_1$ as
\begin{eqnarray}
I_1=\frac{F_1^n(F_{1t}+\dot{x}F_{1x}+\dot{y}F_{1y})}{F_1^nG(t,x)}.
\label{spt01b}
\end{eqnarray}
With this choice one can get
\begin{eqnarray}
&w_1=\frac{1}{(n+1)}F_1^{n+1}(x,t),\quad &dz=F_1^n(x,t)G(t,x)dt,
 \label {spt01c}
\end{eqnarray}
where $n$ is an arbitrary integer. However, to obtain the same independent variable one should multiply
the second integral $I_2$ by the same factor, that is
\begin{eqnarray}
I_2=\frac{F_1^n(F_{2t}+\dot{x}F_{2x}+\dot{y}F_{2y})}{F_1^nG(t,x)}
\label{spt01d}
\end{eqnarray}
which in turn fixes $w_2$ to be in a nonlocal form which contradicts the type-I category.  Thus we conclude that there is no possibility
of getting type-I GST/GLT in infinite sets.

In the following we illustrate the theory with a suitable example.

\subsubsection{Example:2 GST type-I}
To begin with we demonstrate the method of identifying GST of type-I from the integral by considering the two-dimensional Mathews and Lakshmanan
oscillator system \cite{Carinena:04},
\begin{eqnarray}
&&\ddot{x}=\frac{\lambda(\dot{x}^2+\dot{y}^2
+\lambda(x\dot{y}-y\dot{x})^2)x}{(1+\lambda r^2)},\nonumber\\
&&\ddot{y}=\frac{\lambda(\dot{x}^2+\dot{y}^2
+\lambda(x\dot{y}-y\dot{x})^2)y}{(1+\lambda r^2)},
\label{ext01}
\end{eqnarray}
where $r=\sqrt{x^2+y^2}$ and $\lambda$ is an arbitrary parameter. Equation (\ref{ext01}) admits the following two integrals, namely
\begin{eqnarray}
I_1 =\frac{(1+\lambda y^2)\dot{x}-\lambda xy\dot{y}}{\sqrt{1+\lambda r^2}},\quad
I_2=\frac{(1+\lambda x^2)\dot{y}-\lambda xy\dot{x}}{\sqrt{1+\lambda r^2}}.
\label{gfcou10a}
\end{eqnarray}

The integrals (\ref{gfcou10a}) can be rewritten in the form
\begin{eqnarray}
I_1& =\displaystyle{(1+\lambda r^2)\frac{d}{dt}
\bigg(\frac{x}{\sqrt{1+\lambda r^2}}\bigg)}
= \frac {dw_1}{dz_1},\label{gfcou11a}\\
I_2& =\displaystyle{(1+\lambda r^2)\frac{d}{dt}
\bigg(\frac{y}{\sqrt{1+\lambda r^2}}\bigg)}=
 \frac {dw_1}{dz_2},\label{gfcou11b}
\end{eqnarray}
from which we identify the new dependent and independent variables as
\begin{eqnarray}
w_1=\frac{x}{\sqrt{1+\lambda r^2}},\quad w_2=\frac{y}{\sqrt{1+\lambda r^2}},
\quad z=\int \frac{dt}{(1+\lambda r^2)}. \label{gfcou12}
\end{eqnarray}
In terms of these new variables Eq.(\ref{ext01}) can be brought to the form $\frac{d^2w_i}{dz^2}=0,\;\; i=1,2$.

The other two sets can be easily identified by using the expressions (\ref{spgst06}) and (\ref{ceq15}), that is
\begin{eqnarray}
w_1=\frac{x}{y},\quad w_2=\frac{\sqrt{1+\lambda r^2}}{y},
\quad z=\int \frac{1}{y^2}dt \label{gfcou12a}
\end{eqnarray}
and
\begin{eqnarray}
w_1=\frac{\sqrt{1+\lambda r^2}}{x},\quad w_2=\frac{y}{x},
\quad z=\int \frac{1}{x^2}dt. \label{gfcou12b}
\end{eqnarray}
Here also one can check that these new variables modify the Eq.(\ref{ext01}) into free particle equations.  In this example, the new independent variable besides being nonlocal, depends only on $t$, $x$ and $y$.
\subsubsection{Example:3 GLT type-I }
To illustrate this case let us consider a
four dimensional generalized Lotka-Volterra equations  of the form \cite{Brenig:88,Goriely:90,Enders:92}
\begin{eqnarray}
\dot{u}= u^2v+uv, \quad
\dot{v}=v^2+uv^2,\quad
\dot{x}=u, \quad
\dot{y}=yv.
 \label{lscou10a}
\end{eqnarray}
The second order version of the equation (\ref{lscou10a}) may be written
in the form
\begin{eqnarray}
\ddot{x}=\frac{\dot{y}\dot{x}^2+\dot{x}\dot{y}}{y},
\quad \ddot{y}=\frac{\dot{x}\dot{y}^2+2\dot{y}^2}{y}.
\label{lscou10}
\end{eqnarray}
Hereafter we focuss our attention only on the second order version
(\ref{lscou10}).

The first two integrals can be easily identified in the form
\begin{eqnarray}
I_1=\frac{y}{\dot{x}}+y \quad \mbox{and} \quad
I_2 =-\frac{\dot{y}}{y\dot{x}}.
\label{lscou11}
\end{eqnarray}
Rewriting these two first integrals in the form
\begin{eqnarray}
I_1 =\frac{y}{\dot{x}}\frac {d}{dt}(x+t)= \frac {dw_1}{dz_1},\quad
I_2 =\frac{y}{\dot{x}}\frac {d}{dt}(\frac{1}{y})= \frac {dw_2}{dz_2},
\label{lscou11a}
\end{eqnarray}
we identify the following form of linearizing transformations for equation (\ref{lscou10}), namely
\begin{eqnarray}
w_1  = (x+t),\quad w_2=\frac{1}{y}, \quad z =\int\frac{\dot{x}}{y} dt.
 \label{lscou12}
\end{eqnarray}
The remaining two pairs of GLTs are
\begin{eqnarray}
\mbox{(i)} &w_1  = y(x+t),\quad &w_2=y, \quad z =\int y\dot{x} dt\label{lscou12b},\\
\mbox{(ii)} &w_1  = \displaystyle{\frac{1}{(x+t)}},\quad &w_2=\frac{1}{y(x+t)}, \quad z =\int\frac{\dot{x}}{y(x+t)^2} dt.
 \label{lscou12bb}
\end{eqnarray}
The transformations (\ref{lscou12}), (\ref{lscou12b}) and (\ref{lscou12bb}) exhaust the possible generalized linearizing transformations for Eq. (\ref{lscou10}).

In this section we focussed our attention only on the case in which the new dependent variables in the linearized equations are the same. Hereafter we will investigate the situation in which $z_1\neq z_2$.

\section{Type-II Linearizing Transformations ($z_1\neq z_2$)}
Type -II linearizing transformations emerge by allowing the independent variables which appear in the linearized equations to be different, that is $\frac{d^2w_1}{dz_1^2}=0$ and $\frac{d^2w_2}{dz_2^2}=0$. Since the linearized equations are different from each other the
situation is similar to the case of a scalar ODE with the only difference here being that the number of equations to be  derived are two (of course the integrals at hand are also two).

To capture all the transformations one can list out the linearizing transformations that can be obtained from the first integral (which in turn provides the first linearized equation) and from the second integral (which in turn provides the second linearized equation). Now one can pick up any pair $(w_1,z_1)$ obtained through the first integral and combine with any pair
$(w_2,z_2)$ obtained through the second integral which in turn form the linearizing transformation $(w_1, w_2, z_1, z_2)$ for the given two coupled second order nonlinear ODE. This freedom of combination allows us to obtain more number of linearizing transformations in each category as we see below.

Let us start our analysis by considering invertible point transformations.
\subsection{Point transformations of type-II}
The method of linearizing two coupled second order nonlinear ODEs into two distinct free particle equations can be treated as linearizing a scalar ODE. The only difference is that the number of equations to handle are now two. Once this procedure is realized, the remaining part will be in identifying linearizing point transformations from the given two integrals. However, this procedure has already been worked out in Part-I (for the scalar case) \cite{Chand:11}. The result reveals that one can identify the following three pairs of point transformations from a given integral, namely
\begin{eqnarray}
\mbox{(1a)}\; &w_1=F_1,\; z_1=G_1, \;\; &\mbox{(1b)}\; w_{1} = \frac{F_1}{G_1}, \; z_1 = \frac{1}{G_1}\;\;\mbox{and} \nonumber\\
\mbox{(1c)} \; &w_1 = \frac{1}{F_1},\; z_{1} = \frac{G_1}{F_1}.&\label{pt-type21}
\end{eqnarray}

Since we have one more integral at hand the same methodology can be adopted to derive three more linearizing point transformations, namely
\begin{eqnarray}
\mbox{(2a)}\; &w_2=F_2,\; z_2=G_2, \;\; &\mbox{(2b)}\; w_{2} = \frac{F_2}{G_2}, \; z_2 = \frac{1}{G_2}\;\;\mbox{and} \nonumber\\
\mbox{(2c)} \; &w_2 = \frac{1}{F_2},\; z_{2} = \frac{G_2}{F_2}.&\label{pt-type22}
\end{eqnarray}
Interestingly we can pick up any one of the pairs from the first set and any one of the pairs from the second set and linearize the ODE. Obviously this freedom allows one to tabulate the existence of nine pairs of linearizing point transformations of type-II. They are as follows:
\begin{eqnarray}
\mbox{(i)} &w_1=F_1, z_1=G_1; &w_2=F_2, z_2=G_2,\nonumber\\
\mbox{(ii)} &w_1=F_1, z_1=G_1; &w_{2} = \frac{F_2}{G_2}, \;
z_2 = \frac{1}{G_2},\nonumber\\
\mbox{(iii)} &w_1=F_1, z_1=G_1; &w_2 = \frac{1}{F_2},\;
z_{2} = \frac{G_2}{F_2}\nonumber,\\
\mbox{(iv)} &w_{1} = \frac{F_1}{G_1}, \; z_1 = \frac{1}{G_1}; &w_2=F_2, z_2=G_2, \nonumber\\
\mbox{(v)} &w_{1} = \frac{F_1}{G_1}, \; z_1 = \frac{1}{G_1}; &w_{2} = \frac{F_2}{G_2}, \; z_2 = \frac{1}{G_2}, \nonumber\\
\mbox{(vi)} &w_{1} = \frac{F_1}{G_1}, \; z_1 = \frac{1}{G_1}; &w_2 = \frac{1}{F_2},\; z_{2} = \frac{G_2}{F_2}, \nonumber\\
\mbox{(vii)} &w_1 = \frac{1}{F_1},\; z_{1} = \frac{G_1}{F_1};
&w_2=F_2, z_2=G_2,\nonumber\\
\mbox{(viii)} &w_1 = \frac{1}{F_1},\; z_{1} = \frac{G_1}{F_1}; &w_{2} = \frac{F_2}{G_2}, \; z_2 = \frac{1}{G_2}\;\;\mbox{and} \nonumber\\
\mbox{(ix)} &w_1 = \frac{1}{F_1},\; z_{1} = \frac{G_1}{F_1}; &w_2 = \frac{1}{F_2},\; z_{2} = \frac{G_2}{F_2}.
\label{pt-type23}
\end{eqnarray}
We mention here that since the linearizing transformations are of invertible point type each one of the dependent variable may be replaced by an independent variable and vice verse which in turn also yields a linearizing transformation. For example in (\ref{pt-type21}) one can alternatively fix the transformation as
\begin{eqnarray}
\mbox{(1d)}\; &w_1=G_1,\; z_1=F_1, \;\; &\mbox{(1e)}\; w_{1} = \frac{1}{G_1}, \; z_1 = \frac{F_1}{G_1}\;\;\mbox{and} \nonumber\\
\mbox{(1f)} \; &w_1 = \frac{G_1}{F_1},\; z_{1} = \frac{1}{F_1}.&\label{pt-type21aaa}
\end{eqnarray}

This argument is also applicable to the other set (\ref{pt-type22}), that is
\begin{eqnarray}
\mbox{(2da)}\; &w_2=G_2,\; z_2=F_2, \;\; &\mbox{(2e)}\; w_{2} = \frac{1}{G_2}, \; z_2 = \frac{F_2}{G_2}\;\;\mbox{and} \nonumber\\
\mbox{(2f)} \; &w_2 = \frac{G_2}{F_2},\; z_{2} = \frac{1}{F_2}.&\label{pt-type22aaa}
\end{eqnarray}
Combining (\ref{pt-type21aaa}) and (\ref{pt-type22aaa}) suitably one can get the complementary set of
invertible point transformations.

\subsubsection{Example:4 Point transformation}
Let us consider the two dimensional modified Emden type equation
of the form (Carinena \textit{et al.} \cite{Carinena:05})
\begin{eqnarray}
&&\ddot{x}+3k_1x\dot{x}+k_1^2x^3+\lambda_1 x=0,\nonumber\\
&&\ddot{y}+3k_2y\dot{y}+k_2^2y^3+\lambda_2 y=0.
 \label{tcou10}
\end{eqnarray}
Carinena \textit{et al.} \cite{Carinena:05} have proved that equation (\ref{tcou10})
is a super-integrable one, that is, it admits three functionally independent
integrals so that the solution requires only one quadrature.

To derive the linearizing transformation let us concentrate on the first and second integrals, namely
\begin{eqnarray}
I_1=e^{-2\sqrt{-\lambda_1}t}{\bigg(\frac{\dot{x}+k_1x^2+\sqrt{-\lambda_1}x}
{\dot{x}+k_1x^2-\sqrt{-\lambda_1}x}}\bigg), \nonumber\\
I_2=e^{-2\sqrt{-\lambda_2}t}{\bigg(\frac{\dot{y}+k_2y^2+\sqrt{-\lambda_2}y}
{\dot{y}+k_2y^2-\sqrt{-\lambda_2}y}}\bigg). \label{tcou10a}
\end{eqnarray}
Rewriting (\ref{tcou10a}) in the form
\begin{eqnarray}
I_1& =&-\frac{e^{-\sqrt{-\lambda_1}t}k_1x^2}
{{\dot{x}+k_1x^2-\sqrt{-\lambda_1}x}}\left[\frac {d}{dt}
\left((\frac{1}{k_1x}+\frac{1}{\sqrt{-\lambda_1}})e^{-\sqrt{-\lambda_1}t}
\right)\right]= \frac {dw_1}{dz_1},\label{tcou11a}\\
I_2& =&-\frac{e^{-\sqrt{-\lambda_2}t}k_2y^2}
{{\dot{y}+k_2y^2-\sqrt{-\lambda_2}y}}\left[\frac {d}{dt}
\left((\frac{1}{k_2y}+\frac{1}{\sqrt{-\lambda_2}})e^{-\sqrt{-\lambda_2}t}
\right)\right]= \frac {dw_2}{dz_2},\label{tcou11b}
\end{eqnarray}
and identifying the new dependent and independent variables,
\begin{eqnarray}
&&w_1  = F_1=\bigg(\frac{1}{k_1x}+\frac{1}{\sqrt{-\lambda_1}}\bigg)
e^{-\sqrt{-\lambda_1}t},\quad
z_1  = G_1=\bigg(\frac{1}{k_1x}-\frac{1}{\sqrt{-\lambda_1}}\bigg)
e^{\sqrt{-\lambda_1}t},\nonumber\\
&&w_2  = F_2=\bigg(\frac{1}{k_2y}+\frac{1}{\sqrt{-\lambda_2}}\bigg)
e^{-\sqrt{-\lambda_2}t},\quad
z_2  =G_2= \bigg(\frac{1}{k_2y}-\frac{1}{\sqrt{-\lambda_2}}\bigg)
e^{\sqrt{-\lambda_2}t}, \label{tcou12}
\end{eqnarray}
one can linearize Eq. (\ref{tcou10}) as two free particle equations.
Substituting the functions $F_1$, $F_2$, $G_1$ and $G_2$ in (\ref{pt-type23}) one can readily obtain the remaining eight sets of linearizing point transformations for the Eq. (\ref{tcou10}).  The
complementary set of linearizing transformations can also be tabulated from the expressions given
in (\ref{tcou12}).

\subsection{Point-Sundman transformation (PST) and Point-Generalized linearizing transformation (PGT)}
In the previous subsection we considered the case where both the new independent variables are of point type. Let us now fix one of the independent variables to be of point type and other to be nonlocal. Then the resultant transformation should become either a point- Sundman transformation (in the case where the independent variables do not contain  derivative terms) or point-generalized linearizing transformation (when the independent variables do contain derivative terms besides being nonlocal). Since PST differs from PGT only in the independent variable part we consider the method of identifying them from the integral as a single category.

In the present case we have to list out all the linearizing point transformations that can be obtained from an integral and the ST/GLT that can be obtained from the other integral. Now combining a pair of PT with a pair of ST/GLT one can obtain PST/PGLT for a given equation.

We have seen in Part-I \cite{Chand:11} that one can derive three sets of point transformations and infinite number of ST/GLT from a given integral. Similar results hold good here also. The respective forms are
\begin{eqnarray}
\mbox{PT}:\;\; \mbox{(1a)} \;\;  w_1&=&F_1(x,y,t),\; z_1= G_1(t,x,y), \nonumber\\
\mbox{(1b)} \;\; w_1&=&\frac{F_1}{G_1},\;\;\;\;\;\;z_1=\frac{1}{G_1}, \nonumber\\
\mbox{(1c)} \;\;  w_1&=&\frac{G_1}{F_1},\;\;\;\;\;\; z_1=\frac{1}{F_1}, \nonumber\\
\mbox{ST}:\qquad \quad w_2&=&F_2^{n+1},\; dz_2=F_2^nG_2dt,\nonumber\\
\mbox{GLT}:\qquad \quad w_2&=&F_2^{n+1},\; dz_2=F_2^nG_2dt,
\label {PGST-2}
\end{eqnarray}
where $n$ is an arbitrary constant.
From the above, we can formulate the possible linearizing transformations as
\begin{eqnarray}
\mbox{(i)} &w_1=F_1,\; z_1=G_1, &w_2=F_2^{n+1},\; dz_2=F_2^nG_2dt\nonumber\\
\mbox{(ii)} &w_{1} = \frac{F_1}{G_1}, \; z_1 = \frac{1}{G_1}, &w_2=F_2^{n+1},\; dz_2=F_2^nG_2dt \nonumber\\
\mbox{(iii)} &w_1 = \frac{1}{F_1},\; z_{1} = \frac{G_1}{F_1},
&w_2=F_2^{n+1},\; dz_2=F_2^nG_2dt.
\label{pgst-type23}
\end{eqnarray}
Depending upon the nature of $G_2$, the resultant transformation becomes either PST ($G_2=G_2(t,x,y)$) or PGLT $(G_2=G_2(t,x,y,\dot{x},\dot{y}))$.
In the following we illustrate the above procedure by considering an example for each category.

\subsubsection{Example:5 Point-Sundman transformation (PST)}
Let us consider the four dimensional Lotka-Volterra competing population
equation of the form \cite{Brenig:88, Enders:92, Cairo:92, Plank:95}
\begin{eqnarray}
&x_1'=\alpha x_1+\beta  x_1x_2, \quad
&x_2'=\alpha x_2+\gamma  x_1x_2,\nonumber\\
&x_3'=\alpha x_3-\frac{1}{\gamma} x_1x_3, \quad
&x_4'=\alpha x_4-\frac{1}{\beta} x_2x_4,
 \label{gscou10a}
\end{eqnarray}
where prime denotes differentiation with respect to $\tau$ and $\alpha,\beta$
and $\gamma$
are arbitrary parameters.  Using the transformation
$\hat{x_i}=x_ie^{-\alpha\tau},\;i=1,...4,$ and
$t=e^{\alpha\tau}$ equation (\ref{gscou10a}) can be transformed to the
form
\begin{eqnarray}
\dot{\hat{x}}_1=\beta x_1x_2, \quad
\dot{\hat{x}}_2=\gamma x_1x_2 ,\quad
\dot{\hat{x}}_3=\frac{1}{\gamma} x_1x_3, \quad
\dot{\hat{x}}_4=\frac{1}{\beta} x_2x_4.
 \label{gscou10b}
\end{eqnarray}
For the present purpose let us rewrite equation (\ref{gscou10b}) as a system of two coupled second order nonlinear ODEs of the form
\begin{eqnarray}
\ddot{x}=\frac{\dot{x}^2}{x}+\frac{\dot{x}\dot{y}}{y},
\quad \ddot{y}=\frac{\dot{y}^2}{y}+\frac{\dot{x}\dot{y}}{x}, \label{gscou10}
\end{eqnarray}
where $x=\hat{x}_3$ and $y=\hat{x}_4$. Eq. (\ref{gscou10}) posseses two integrals $
I_1 =\frac{\dot{x}}{\dot{y}}$ and $I_2=\frac{\dot{y}}{xy}$. The point transformation that can be extracted from $I_1$ is
\begin{eqnarray}
w_1=x, \quad z_1=y. \label{gscou12}
\end{eqnarray}
The ST that can be extracted from $I_2$ is
\begin{eqnarray}
w_2=\log(y),\quad z_2=\int x dt. \label{gscou12aa}
\end{eqnarray}
Now utilizing the expressions given in (\ref{pgst-type23}) one can enlist the linearizing transformations for (\ref{gscou10}) which in turn read as
\begin{eqnarray}
\mbox{(i)} &w_1=x, \; z_1=y,  &w_2=\log(y)^{n+1},\; dz_2=x\log(y)^ndt\nonumber\\
\mbox{(ii)} &w_{1} = \frac{x}{y}, \; z_1 = \frac{1}{y}, &w_2=\log(y)^{n+1},\; dz_2=x\log(y)^ndt \nonumber\\
\mbox{(iii)} &w_1 = \frac{1}{x},\; z_{1} = \frac{y}{x},
&w_2=\log(y)^{n+1},\; dz_2=x\log(y)^ndt.
\label{pgst-type23ex}
\end{eqnarray}

In the above example the second independent variable, $z_2$, is only a function $t$, $x$ and $y$. Now we consider an example in which the second variable is a function of $t$, $x$, $y$, $\dot{x}$ and $\dot{y}$.
\subsubsection{Example:6 Point-generalized linearizing transformation}
To demonstrate this category we consider a
variant of the two-dimensional Mathews and Lakshmanan equation of the form
\begin{eqnarray}
&&\ddot{x}=\frac{\lambda(\dot{x}^2+\dot{y}^2
+2 \lambda(\dot{y}-\dot{x})^2)-\alpha^2}{(1+2\lambda (x+y))},\nonumber\\
&&\ddot{y}=\frac{\lambda(\dot{x}^2+\dot{y}^2
+2 \lambda(\dot{y}-\dot{x})^2)-\alpha^2}{(1+2\lambda (x+y))}
\label{fglt01}
\end{eqnarray}
which admits two constants of motion, namely
\begin{eqnarray}
I_1 =\dot{x}-\dot{y},\quad
I_2=\frac{\alpha^2-\lambda((1+2\lambda)(\dot{y}-\dot{x})^2+2\dot{x}\dot{y})}
{(1+2\lambda (x+y))}.
\label{fglt02}
\end{eqnarray}
From $I_1$ and $I_2$ we identify the following transformation,
\begin{eqnarray}
&w_1=(x-y),\quad &w_2=\log(1+2\lambda (x+y)),\nonumber\\
&z_1=t,\quad &z_2=\int\frac{2\lambda (\dot{x}+\dot{y})}
{\alpha^2-\lambda((1+2\lambda)(\dot{y}-\dot{x})^2+2\dot{x}\dot{y})}dt.
\label{fglt04}
\end{eqnarray}
Eq. (\ref{pgst-type23}) allows us to obtain the following linearizing transformations for (\ref{fglt01}) as
\begin{eqnarray}
\mbox{(i)} &w_1=(x-y),\; z_1=t, &w_2=\log(1+2\lambda (x+y))^{n+1},\;
\nonumber\\ &&dz_2=\frac{2\lambda (\dot{x}+\dot{y})\log(1+2\lambda (x+y))^n}
{h(x)}dt,\nonumber\\
\mbox{(ii)} &w_{1} = \frac{(x-y)}{t}, \; z_1 = \frac{1}{t}, &w_2=\log(1+2\lambda (x+y))^{n+1},\; \nonumber\\ &&dz_2=\frac{2\lambda (\dot{x}+\dot{y})\log(1+2\lambda (x+y))^n}
{h(x)}dt\;\;\mbox{and} \nonumber\\
\mbox{(iii)} &w_1 = \frac{t}{(x-y)},\; z_{1} = \frac{t}{(x-y)},
&w_2=\log(1+2\lambda (x+y))^{n+1},\; \nonumber\\ && dz_2=\frac{2\lambda (\dot{x}+\dot{y})\log(1+2\lambda (x+y))^n}
{h(x)}dt,
\label{pgst-type23ex2}
\end{eqnarray}
where $h(x)=\alpha^2-\lambda((1+2\lambda)(\dot{y}-\dot{x})^2+2\dot{x}\dot{y})$. One may note that
in all the above transformations
the second independent variable is a function of $t$, $x$, $y$, $\dot{x}$ and $\dot{y}$.

\subsection{Sundman transformation of type-II (ST-II), Sundman-generalized linearizing transformation (SGLT) and generalized linearizing transformation of type-II (GLT-II)}
Finally we group the cases in which both the independent variables are in nonlocal form. The possible forms entering into this category are (i) ST type-II (both the independent variables are only functions of $t$, $x$ and $y$ alone) (ii) GLT type-II (both the independent variables are  functions of $t$, $x$, $y$, $\dot{x}$ and $\dot{y}$)  (iii) SGLT (only one of the independent variables contains derivative terms).
Since one can identify infinite number of GLT/ST's from a given integral the above types of linearizing transformations notably will be larger in number than the rest.

The possible forms of linearizing transformations are
\begin{eqnarray}
&w_1=F_1^{m+1},\quad &dz_1=F_1^mG_1dt, \quad m\neq-1\nonumber\\
&w_2=F_2^{n+1},\quad &dz_2=F_2^nG_2dt, \quad n\neq-1
\label {st209}
\end{eqnarray}
and
\begin{eqnarray}
&w_1=\log{F_1},\quad &dz_1=\frac{G_1}{F_1}dt, \quad \;\;\; m=-1\nonumber\\
&w_2=\log{F_2},\quad &dz_2=\frac{G_2}{F_2}dt, \quad \;\;\;n=-1\label {st210}
\end{eqnarray}
where $m$ and $n$ are some arbitrary numbers.

\subsubsection{Example:7 Sundman transformation of type-II (ST-II)}
To begin with let us consider an example for ST-II.
To illustrate the ST-II we consider the same equation, (\ref{gscou10}),
which we considered to illustrate the PST type.
However, for the present analysis we take the integrals in the form
$I_1 =\frac{\dot{x}}{xy}$ and $I_2=\frac{\dot{y}}{xy}$.
Rewriting these two integrals as
\begin{eqnarray}
I_1 =\frac{1}{y}\frac {d}{dt}(\log(x))= \frac {dw_1}{dz_1}\quad \mbox{and}\quad
I_2 =\frac{1}{xy}\frac {d}{dt}(y)= \frac {dw_2}{dz_2},
\label{gtcou11}
\end{eqnarray}
we identify the linearizing transformations in a more general form
\begin{eqnarray}
w_1=\log(x),\quad w_2=y,\quad z_1=\int y dt,\quad z_2=\int xy dt.
\label{gtcou12}
\end{eqnarray}
Other transformations are (vide Eqs.(\ref{st209}))
\begin{eqnarray}
&w_1=(\log(x))^{m+1},\quad &dz_1=(\log(x))^mydt, \quad m\neq-1\nonumber\\
&w_2=y^{n+1},\quad &dz_2=y^{n+1}xdt, \quad n\neq-1
\label {1glt209GGT2a}
\end{eqnarray}
and
\begin{eqnarray}
&w_1=\log(\log(x)),\quad &dz_1=\frac{y}{\log(x)}dt, \quad \;\;\; m=-1\nonumber\\
&w_2=\log{y},\quad &dz_2=xdt, \quad \;\;\; n=-1\label {1glt210GGT2ab}
\end{eqnarray}
where $m$ and $n$ are arbitrary constants.

\subsubsection{Example:8 Sundman-Generalized linearizing
transformation}
To illustrate this type of linearizing transformations let us consider an
equation which is similar to Example 1, that is,
\begin{eqnarray}
\ddot{x}+\dot{x}^2+\dot{x}\dot{y}=0,\qquad
\ddot{y}-\dot{x}\dot{y}=0
\label{sgcou10}
\end{eqnarray}
which admits the first integrals in the form
$I_1 =\dot{x}e^{(x+y)}$ and $I_2=\dot{x}\dot{y}e^{y}$.
Rewriting the first integrals as
$I_1 =e^{y}\frac {d}{dt}(e^{x})$ and
$I_2 =\dot{x}\frac {d}{dt}(e^{y})$ we identify the linearizing
transformations of the form
\begin{eqnarray}
w_1=e^{x},\quad w_2=e^{y},\quad z_1=\int e^{-y} dt,\quad z_2=\int
\frac{1}{\dot{x}} dt.
\label{sgcou12}
\end{eqnarray}
Other transformations are
\begin{eqnarray}
&w_1=(e^{x})^{m+1},\quad &dz_1=\frac{\dot{y}}{\dot{x}}e^{mx}dt, \quad m\neq-1\nonumber\\
&w_2=e^{(n+1)y},\quad &dz_2=\frac{e^{ny}}{\dot{x}}dt, \quad n\neq-1
\label {2glt209GGT2a}
\end{eqnarray}
and
\begin{eqnarray}
&w_1=x,\quad &dz_1=\frac{\dot{y}}{\dot{x}}e^{-x}dt, \quad m=-1\nonumber\\
&w_2=y,\quad &dz_2=\frac{1}{\dot{x}}e^{-y}dt, \quad n=-1\label {2glt210GGT2ab}
\end{eqnarray}
where $m$ and $n$ are arbitrary constants.
\subsubsection{Example:9 Generalized linearizing transformation of type-II}
To understand the generalized linearizing transformations let us start
with the following system of four coupled first-order ODEs
of the form \cite{Brenig:88, Enders:92,
Cairo:92, Plank:95,Golubchik:00, Sakovich:01}
\begin{eqnarray}
&\dot{x}_1=\alpha_1 x_1^2+\alpha_2 x_1x_2, \; &
\dot{x}_2=\beta_1 x_2^2+\beta_2 x_1x_2,\nonumber\\
&\dot{x}_3=\gamma_1x_1x_3+\gamma_2x_2x_3, \;&
\dot{x}_4=\delta_1x_1x_4+\delta_2x_2x_4,
 \label{case901}
\end{eqnarray}
where $\alpha_i,\beta_i,\gamma_i$ and  $\delta_i$, $i=1,2,$
are arbitrary parameters. Introducing the new variables
$\hat{x}_3=x_3^{\frac{\gamma_1}{\omega_1}}x_4^{-\frac{\gamma_2}{\omega_2}}$ and
$\hat{x}_4=x_3^{-\frac{\delta_1}{\omega_1}}x_4^{\frac{\delta_2}{\omega_2}}$,
where $\omega_i$'s, $i=1,2,$ are arbitrary parameters \cite{Sakovich:01} in
equation (\ref{case901}) we get
\begin{eqnarray}
\dot{x}_1=\alpha_1 x_1^2+\alpha_2 x_1x_2, \;
\dot{x}_2=\beta_1 x_2^2+\beta_2 x_1x_2, \;
\dot{\hat{x}}_3=\hat{\gamma_1} x_1\hat{x}_3, \;
\dot{\hat{x}}_4=\hat{\delta_1} x_2\hat{x}_4,
 \label{case902}
\end{eqnarray}
where $\hat{\gamma_1}=\frac{(\gamma_1\delta_2-\gamma_2\delta_1)}{\omega_2}$ and
$\hat{\delta_1}=\frac{(\gamma_1\delta_2-\gamma_2\delta_1)}{\omega_1}$.
Equation (\ref{case902}) may be written as a set of coupled second order ODES of the
form
\begin{eqnarray}
\ddot{x}=(1-\frac{\alpha_1}{\hat{\gamma_1}})\frac{\dot{x}^2}{x}
-\frac{\alpha_2}{\hat{\delta_1}}\frac{\dot{x}\dot{y}}{y},
\quad \ddot{y}=(1-\frac{\beta_1}{\hat{\delta_1}})\frac{\dot{y}^2}{y}
-\frac{\beta_2}{\hat{\gamma_1}}\frac{\dot{x}\dot{y}}{x}, \label{case903}
\end{eqnarray}
where $\hat{x}_3=x$ and $\hat{x}_4=y$. One may note that equations
(\ref{case903})
and (\ref{gscou10}) coincide under the parametric restrictions
$\alpha_1=0,\;\beta_1=0,$ $\hat{\gamma_1}=\frac{1}{\beta_2}$ and
$\hat{\delta_1}=\frac{1}{\alpha_2}$. However, for the present illustration we
consider another linearizable equation in this class by restricting the
parameters to
$\hat{\gamma_1}=\frac{3}{2}\alpha_1=3\beta_2$ and $\hat{\delta_1}
=3\alpha_2=\frac{3}{5}\beta_1$. Under this choice, equation (\ref{case903})
assumes the form
\begin{eqnarray}
\ddot{x}=\frac{\dot{x}^2}{3x}-\frac{\dot{x}\dot{y}}{3y},
\quad \ddot{y}=-\frac{2\dot{y}^2}{3y}-\frac{\dot{x}\dot{y}}{3x}.
\label{case904}
\end{eqnarray}
The associated first integrals turn out to be
\begin{eqnarray}
I_1=\frac{\dot{x}^2}{x\dot{y}},\qquad
I_2 =2y\dot{y}\dot{x}.
\label{case905}
\end{eqnarray}
Rewriting (\ref{case905}) in the form
$I_1 =-\frac{\dot{x}}{\dot{y}} \frac{d}{dt}(\log(x))$ and
$I_2 =\dot{x}\frac {d}{dt}y^2$ and identifying the new variables we get
\begin{eqnarray}
w_1  = \log(x),\;w_2=y^2,\; z_1  = \int \frac{\dot{y}}{\dot{x}}dt,\;
z_2=\int \frac{1}{\dot{x}}dt.
 \label{case906}
\end{eqnarray}
Other transformations are
\begin{eqnarray}
&w_1=\log(x)^{m+1},\quad &dz_1=\frac{\dot{y}}{\dot{x}}\log(x)^mdt, \quad m\neq-1\nonumber\\
&w_2=y^{2(n+1)},\quad &dz_2=\frac{y^{2n}}{\dot{x}}dt, \quad n\neq-1
\label {glt209GT2a}
\end{eqnarray}
and
\begin{eqnarray}
&w_1=\log{\log(x)},\quad &dz_1=\frac{\dot{y}}{\dot{x}\log(x)}dt, \quad \;\;\; m=-1\nonumber\\
&w_2=\log{y^2},\quad &dz_2=\frac{1}{\dot{x}y^2}dt, \quad \;\;\; n=-1\label {glt210GT2ab}
\end{eqnarray}
where $m$ and $n$ are arbitrary constants.

\section{Conclusion}
In this paper, we have extended the method of deriving the maximal number of linearizing transformations from a knowledge of the integrals for the two coupled second order nonlinear ODEs. The procedure proposed here is simple and straightforward and yields not only the expected point, Sundman and generalized linearizing transformations but also gives hybrid ones, including point-Sundman, point-GLT and ST-GLT. Interestingly we have demonstrated that whatever be the form or type of transformations they can be extracted in maximum from the integrals admitted by the equation. We have proved the applicability of the algorithm by considering a number of examples. To our knowledge no single algorithm has been shown to yield this much variety of linearizing transformations. In that sense, we believe that we have established a stand alone method to derive LTs in the theory of linearization of nonlinear ODEs.

\section*{Acknowledgments}
The work of VKC and ML is supported by a Department of Science and Technology (DST), Government of India, IRHPA research project. ML is also supported by a DAE Raja Ramanna Fellowship and a DST -- Ramanna program. The work of MS forms part of a research project sponsored by the DST.

%\section*{References}

\end{document}